\documentclass[twocolumn,showpacs,prl,floatfix,amssymb,superscriptaddress]{revtex4}
\usepackage{graphicx}  
\usepackage{dcolumn}  
\usepackage{bm}  
\usepackage[colorlinks=true]{hyperref}  
\usepackage{natbib}
\usepackage[normalem]{ulem}
\usepackage{color}
\usepackage{epstopdf}

\newcommand{\hidden}[1]{}

\begin{document}

\title{Electron spin relaxation as evidence of excitons in a two dimensional electron-hole plasma}

\author{S. Oertel}

\author{S. Kunz}
\affiliation{Institute for Solid State Physics, University Hannover, 30167 Hannover, Germany}
\author{D. Schuh}
\affiliation{Institute for Experimental and Applied Physics, University Regensburg, 93040 Regensburg, Germany}
\author{W. Wegscheider}
\affiliation{Solid State Physics Laboratory, ETH Z{\"u}rich, 8093
Z{\"u}rich, Switzerland}

\author{J. H{\"u}bner}
\author{M. Oestreich}
\affiliation{Institute for Solid State Physics, University Hannover, 30167 Hannover, Germany}
\date{\today}


\begin{abstract}

We exploit the influence of the Coulomb interaction between
electrons and holes on the electron spin relaxation in a (110)-GaAs quantum
well to unveil excitonic signatures within the many particle
electron-hole system. The temperature dependent time- and
polarization-resolved photoluminescence measurements span five
decades of carrier density, comprise the transition from localized
excitons over quasi free excitons to an electron-hole plasma, and
reveal strong excitonic signatures even at relatively high
densities and temperatures.
\end{abstract}

\maketitle

Shortly after the big bang, the hot gas of negatively charged
electrons and positively charged protons cooled to temperatures
below the Rydberg energy and formed a new state of matter, called
hydrogen atoms. 13.7 billion years later, researchers shoot short
laser pulses on direct semiconductors at liquid helium temperature
creating a hot gas of negatively charged electrons in the
conduction and positively charged holes in the valence band. The
hot carrier gas cools with time by emission of optical and
acoustical phonons and forms excitons, i.e., hydrogen like quasi-particles. These quasi-particles have in bulk GaAs a binding
energy of 5~meV and strongly influence the photoluminescence (PL)
spectrum. Consequently, the PL emission at the
exciton transition energy has been used as an indicator for the
existence of excitons\cite{Feldmann1987} -- as the emission and absorption
lines of hydrogen are used in astronomy. However, S. Koch and
coworkers demonstrated within the framework of many body
semiconductor Bloch equations that distinct excitonic like emission
lines from an electron-hole plasma do not proof the existence of
excitons but can be  explained equally by a Coulomb correlated
electron-hole plasma.\cite{Koch2006}

Manyfold interband pump-probe absorption\cite{Knox1985}, reflection\cite{Malinowski2000}, and four wave mixing experiments\cite{Noll1990}, high resolution time-resolved PL\cite{Szczytko2004}, and experimentally very demanding quasi-particle THz spectroscopy\cite{Kaindl2003} have been carried out to understand the many-body and quantum-optical character and the diligent interaction of excitons in an electron-hole plasma but besides fifty years of intense research the exciton quest remains a fascinating and active field. The challenge in the interpretation of most interband experiments concerning the incoherent exciton population results from the fact
that the interaction of classical light and matter is induced by
optical polarization and not directly by incoherent population.
The challenge in the interpretation of time-resolved PL
experiments results generally from the finite excitation density
and that thereby the PL does not result from a two but a many
particle interaction which strongly depends on the
frequency-dependent strength of light-matter interaction and only
to some extend on exciton population. Especially THz experiments
support that excitonic like PL lines exist without the population
of K=0 excitons.\cite{Kaindl2009} In this letter, we present an entirely different
experimental approach to study the existence of excitons in a
two-dimensional electron-hole plasma and utilize the electron-hole
spin interaction as exciton marker. In general, the spin dynamics
in GaAs quantum wells (QWs) is not an appropriate indicator for excitons
since the spin of free holes dephases rapidly within the hole
momentum scattering time due to the mixing of heavy hole (HH) and light hole (LH)
and the electron spin dephases in (001)-GaAs-QWs rapidly due to
the Dyakonov-Perel (DP) spin relaxation mechanism, i.e., both electron
and hole spin relaxation do not significantly depend on exciton
formation. Fortunately, the DP spin relaxation of
electrons can be easily suppressed by the symmetry of the QW.
Several groups have demonstrated that the DP mechanism vanishes in
(110)-GaAs-QWs for electron spins pointing parallel or
antiparallel to the growth direction and that the Elliott-Yafet
and the intersubband spin relaxation (ISR) mechanism are inefficient in
thin QWs.\cite{Lombez2007} In the case of thin (110)-QWs, the only remaining
significant electron spin relaxation mechanisms are the
Bir-Aronov-Pikus (BAP) and the exciton exchange spin relaxation.\cite{Wu2010}
Very recently, Zhou and Wu have calculated by sophisticated
semiconductor spin Bloch equations that the BAP mechanism is at
moderate carrier densities surprisingly inefficient in
two-dimensional samples,\cite{Zhou2008} i.e., efficient electron spin relaxation
results in (110)-GaAs-QWs predominantly from excitonic exchange
interaction and is thereby a clear measure for the existence of
excitons in an electron-hole plasma.

\begin{figure}[tbh]
    \centering
        \includegraphics[width=1.0\columnwidth]{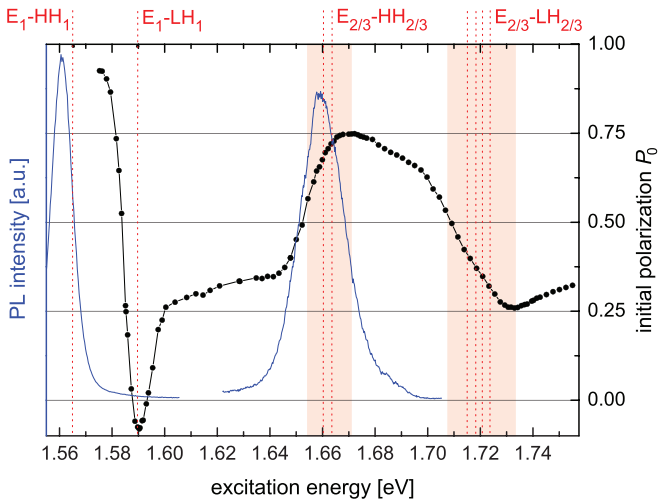}
    \caption{(color online) Initial polarization of the
    PL of the thick QW in dependence on
    the excitation energy (black points). The solid blue lines are the
    PL of the thick QW (left side) and of the thin QWs (middle)
    with different arbitrary units. The dashed red lines are the
    calculated HH and LH transitions of the
    complete structure (the indices correspond to the three lowest energy levels). The light red areas show the calculated
    broadened transition energies due to monolayer fluctuations in
    the thin QWs.} \label{fig_energy_dependence}
\end{figure}

In this letter, we study the existence of excitons in an electron-hole
plasma by measuring the density and temperature dependence of
the electron spin relaxation time in an undoped 9~nm (110)-QW by time- and polarization resolved PL. The sample is designed
as a novel, optical, non-resonant, high efficient spin injection
structure by sandwiching the 9~nm QW between two 4~nm QWs
separated by 3~nm Al$_{0.36}$Ga$_{0.64}$As barriers. We excite the
sample by circularly polarized, 1.5~ps-pulses from a mode-locked
Ti:Sapphire laser with at repetition rate of 80~MHz and a focus
diameter of 100~$\mu$m. The circular polarization yields according
to the optical selection rules spin polarized electrons whose spin
polarization are in turn detected by the circular polarization of
the PL. The PL is detected by a streak-camera system with a
spectral and temporal resolution of 3~meV and 6~ps, respectively.
The right and left circular polarization of the PL ($I_-$ and
$I_+$) are sequentially detected by a computerized liquid crystal
retarder and a linear polarizer.

First, we characterize the sample concerning the optical electron
spin injection. Figure~\ref{fig_energy_dependence} shows the
initial degree of circular polarization $P_0=(I_--I_+)/(I_-+I_+)$
of the thick QW PL in dependence on excitation
wavelength for a lattice temperature of 10~K. The initial degree
of polarization is extracted from the time-resolved degree of
polarization by extrapolation to time equal zero. This method
avoids any influence of laser stray light on the degree of PL
polarization. The figure clearly shows nearly 100~\% polarization
for quasi-resonant (E$_1$-HH$_1$) and about 30~\% polarization for non-resonant
excitation of the wide QW and about 75~\%
polarization for resonant excitation (E$_{2/3}$-HH$_{2/3}$) of the thin QWs.
\cite{note1} The latter proves that the electrons' spin is
conserved while tunnelling from the thin into the thick QW. The
carrier tunnelling time is $\le 3$~ps as measured from the thin QW
PL decay time at non-resonant excitation. The electron spin polarization
is not 100\% for resonant excitation of the HH
transition of the thin QWs since we also excite the continuum of
the wide QW.
The degree of the initial PL polarization remains constant at 75~\%
for temperatures up to 70~K and decreases for higher temperatures
exponentially to 22~\% at 200~K. The
degree of polarization decreases with increasing temperature since
the LH contribution rises and, in the case of resonant
excitation of the thin QWs, due to the decrease of the excitonic
enhancement of the HH transition. We want to point out that such a
high efficient optical spin injection structures not only
significantly increases the sensitivity of our spin relaxation
measurements but is also useful for many other experiments like for
optically pumped spin VCSELs or polarization dependent spin
relaxation experiments. In the following, all measurements are
carried out for resonant excitation of the HH transition of the
thin QWs.

In the next step, we distinguish between localized and
non-localized carriers by the spectral width and lifetime of the
PL and by spin quantum beat spectroscopy. The PL spectrum is at 10~K inhomogeneously broadened due to
QW width fluctuations with a full width at half maximum (FWHM) of
about 7.5~meV for excitation densities $\le 1 \times
10^{11}$~cm$^{-2}$. The FWHM of the PL is larger than in
comparable (001)-GaAs-QWs due to the more complex growth dynamics
of (110)-QWs. The PL maximum remains at constant emission energy for densities
below $5 \times 10^{9}$~cm$^{-2}$, increases sharply by 2~meV with
increasing density, and remains constant again for densities between $2
\times 10^{10}$~cm$^{-2}$ and  $2 \times 10^{11}$~cm$^{-2}$. This
indicates a transition from localized to unlocalized electrons
and an electron disorder localization potential of 2~meV. The energy shift
of the PL maximum with density vanishes at temperatures of about
20~K to 30~K confirming the potential depth of 2~meV. The PL
rise time at 10~K is faster than our time resolution and the decay is
purely monoexponential for low excitation densities
showing that the excited carriers are rapidly trapped in the
localization potential. In contrast, the PL transient shows at the
same temperature but an excitation density of $2\times
10^{10}$~cm$^{-2}$ a rise time of about 100~ps which becomes more
pronounced and long-lasting with increasing density. The slow rise
time vanishes at a lattice temperature of 50~K and is an
unambiguous signature for cooling of free carriers, i.e., we
observe at 10~K with increasing density a transition from
localized to free electrons and for temperatures $\ge 50$~K only
free electrons. We know that the 2~meV is the electron trapping
potential from spin quantum beat experiments whereat the measured
electron Land\'e g-factor
$g_e$ depends at low temperatures and low densities on the PL
energy and increases linearly from -0.119 at 1.5545~eV to -0.1125
at 1.565~eV. In contrast, $g_e$ is independent of PL energy at
$T\ge30$~K, i.e., the electrons are
delocalized. Accordingly, we study in the following a
two-dimensional system where the electrons and holes experience a
2~meV and a $\approx 7$~meV localization potential, respectively.

\begin{figure}[tb]
    \centering
        \includegraphics[width=1.0\columnwidth]{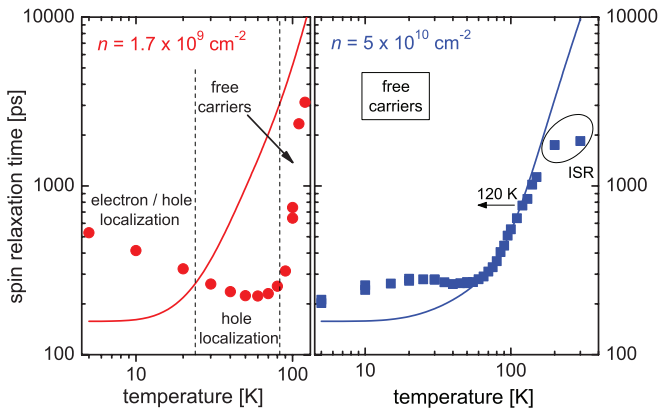}
    \caption{(color online) Measured temperature dependence of
    $\tau_s$ for excitation densities $n =
    1.7 \times 10^{9}$~cm$^{-2}$ (left, red dots) and $5 \times
    10^{10}$~cm$^{-2}$ (right, blue squares) and calculated $\tau_s$
    by excitonic exchange interaction for a plasma of free electrons,
     holes, and excitons (solid lines). The dashed lines in
    the left figure correspond to the thermal energies of the
    localization potentials.} \label{fig:TempDepTauBzero}
\end{figure}

With this background in mind, we study first the temperature
dependence of the electron spin relaxation time $\tau_s$.
Fig.~\ref{fig:TempDepTauBzero} depicts $\tau_s$ in dependence on
temperature for low excitation density $n=1.7\times 10^9$cm$^{-2}$
(left) and high excitation density $n=5\times 10^{10}$cm$^{-2}$
(right). In the low density case, $\tau_s$ decreases gradually
from 500~ps at 5~K to 200~ps at 60~K. This decrease reflects the
transition from localized carriers -- like in quantum dots -- to
free electrons and weakly bound holes. At $T\ge80$~K, also the
holes become delocalized and $\tau_s$ increases sharply by one
order of magnitude between 80~K and 120~K.
We will
show in the next paragraph that these long $\tau_s$ result from
the very low exciton fraction $f_X\equiv n_X/(n_X+n_{e-h})$ in the
low density electron-hole plasma whereat $n_X$ is the exciton
density and $n_{e-h}$ the electron-hole density.

In the high density case, we observe for all temperatures mainly
delocalized electrons and holes. The electron spin relaxation time
is in this case nearly constant between 5~K and 60~K and increases
by about 600~\%\ between 60~K and room temperature. This measured
increase of $\tau_s$ is in quantitative agreement with the
calculated electron spin relaxation by excitons including exciton
ionization by optical phonons (solid line). A comparison with the
low density case shows that $\tau_s$ is at 120~K significantly
shorter in the high density case. This fact might be surprising at
first since a higher density implies more efficient motional
narrowing but the following calculations show that the shorter
$\tau_s$ is directly linked with a significantly higher exciton
fraction.

The spin relaxation of electrons in a plasma of free electrons,
free holes, and free excitons is dominated in (110)-QWs at
$T<200$~K by spin relaxation due to excitons. The exciton fraction
in a 2D electron gas in the Boltzmann limit depends on the carrier
density and the temperature and is described by the so called Saha
equation
\begin{equation}\label{equ:Saha}
  \frac{(n_{e-h})^2}{n_X}= \frac{k_B T}{2\pi \hbar^2}\mu e^{-E_0/k_BT},
\end{equation}
where $\mu=0.061 m_0$ is the reduced mass and $E_0=8$~meV the
exciton binding energy. The resulting exciton fraction at 120~K is
for example 3.5~\% at a density of $2\times 10^9$~cm$^{-2}$ and
37~\% at $6\times 10^{10}$~cm$^{-2}$. The spin relaxation time of
pure excitons reads $\tau_s^{exc}=(\Omega^2 \tau_p)^{-1}$
whereat $\Omega$ is dominated by the long-range exchange
interaction which is according to calculations by Maialle et al.\cite{Maialle1993}
$\Omega = 69.85$~GHz for a 9~nm GaAs-QW and $\tau_p$
is the excitons center of mass momentum scattering time. We
calculate $\tau_p$ by calculating the LO-phonon scattering rate
according to Ref.~\cite{Ridley1988} and by introducing phenomenologically a
constant scattering rate of 0.77~ps$^{-1}$ to account for surface
roughness scattering and all other residual scattering mechansims.
The resulting electron spin relaxation time
in an electron, hole, exciton plasma is the exciton spin
relaxation time weighted by the exciton fraction:
$\tau_s=\tau_s^{exc}/f_X$. The solid lines in
Fig.~\ref{fig:TempDepTauBzero} depict the calculated $\tau_s$ for
free carriers. For the high density case and $T<200$~K, the
calculations are in very good agreement with the experiment while
for $T\ge200$~K, the measured $\tau_s$ is shorter than calculated
for two reasons. Firstly, the electrons scatter at high
temperatures efficiently into higher subbands and ISR becomes important. Secondly, we observe a
significant decrease of the PL lifetime with increasing
temperature for $T>200$~K, i.e., electrons jump out of the QW,
loose their spin orientation due to the very efficient DP
mechanism in the barrier, and a fraction of the depolarized
electrons falls back into the QW reducing the average electron
spin orientation in the QW. For low densities and $T\le100$~K,
the discrepancy between experiment and theory is tremendous. This
is also not surprising since localization dominates the spin
relaxation process. More important, for $T>100$~K all carriers
become delocalized and the measured $\tau_s$ approaches theory.
For $T>120$~K, reliable measurements become difficult since
$\tau_s$ becomes much longer than the PL lifetime and the laser
repetition rate. Please note, that the measured $\tau_s$ at 120~K
is significantly longer in the low density case as
predicted by the Saha equation.

We have also measured the inplane spin relaxation time by applying
a transverse magnetic field. The inplane spin relaxation time is
dominated by the DP spin relaxation and directly yields the
electron momentum scattering time $\tau_p^e$. This extracted $\tau_p^e$
is at temperatures below 200~K much shorter than the calculated LO-phonon
scattering time and thus related to electron-electron scattering.
Nevertheless, the LO-phonon scattering time dominates the excitonic
$\tau_s$ confirming the theoretical prediction that electron-electron scattering does not yield motional
narrowing in the case of excitonic electron spin relaxation.

\begin{figure}[tb]
    \centering
        \includegraphics[width=1.0\columnwidth]{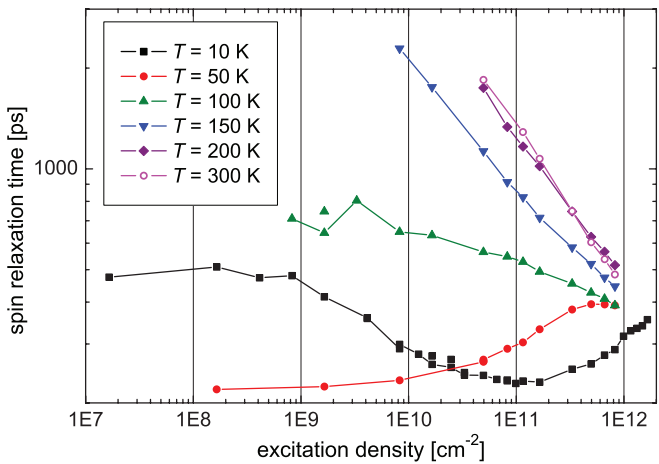}
    \caption{(color online) Spin relaxation time versus excitation densities for
    different temperatures ranging from 10~to 300~K (the lines are guides to the eye).}
    \label{fig:PowerTempDepTauBzero}
\end{figure}

Next, we study the density dependence of $\tau_s$ for temperatures
between 50~K and 300~K (see Fig.~\ref{fig:PowerTempDepTauBzero}).
At $T\ge50$~K, the electron spin decays in good approximation monoexponential
for all carrier densities which is consistent with the picture of
delocalized electrons. The slight deviation from a monoexponential
decay results from the change of carrier density in the
measurement window due to radiative recombination. Most
interestingly, the measured electron spin relaxation time is for
densities $\ll 10^{12}$~cm$^{-2}$ \emph{not} consistent with the
calculated spin relaxation by the BAP mechanism which is at least
one order of magnitude less efficient for densities $\le
10^{11}$~cm$^{-2}$ (see Ref.~\cite{Zhou2008}). The times are also
not consistent with electron spin relaxation times where holes are
absent since previous spin noise spectroscopy measurements in n-doped (110)-QWs with an
electron density of $1.1\times 10^{11}$~cm$^{-2}$ yield by two
orders of magnitude longer electron spin relaxation
times.\cite{Muller2008} In fact, the measured spin relaxation times
can only be explained by the existence of excitons and efficient
electron spin relaxation by exciton exchange interaction, i.e.,
the density dependent measurements confirm that $\tau_s$ is really
a measure of the excitonic influence in an electron hole plasma.
For densities between $10^{11}$~cm$^{-2}$ and $10^{12}$~cm$^{-2}$,
$\tau_s$ converges with increasing carrier density to
approximately the same $\tau_s$ for all temperatures between 50~K
and 300~K. This measured $\tau_s$ of $\approx 450$~ps at a carrier
density of $10^{12}$~cm$^{-2}$ is in good agreement with
sophisticated calculations by semiconductor Bloch equations for
(001)-QWs. Thereby, the calculations by Zhou and Wu prove that the
electron spin relaxation is dominated at very high densities not
by excitonic spin relaxation but by the BAP mechanism. Our
experiments also confirm the predictions by Zhou and Wu that the BAP
mechanism is surprisingly temperature insensitive.

Last, we study the density dependence of $\tau_s$ at $T=10$~K where
localization and Pauli blockade plays an important role. At this
lattice temperature, the spin relaxation is monoexponential at
low densities with $\tau_s\approx 500$~ps and becomes
biexponential at densities $\ge 2\times 10^9$~cm$^{-2}$ whereat
Fig.~\ref{fig:PowerTempDepTauBzero} depicts the initial fast
decay. A clear transition from mono- to biexponential spin
relaxation appears at the same density whereat the PL shows a
transition from localized to free electrons, i.e., this initial
fast $\tau_s$ results from free electrons. The electron spin
relaxation time decreases with decreasing localization (increasing
density) since the spin relaxation time of localized holes is extremely long in
comparison to free holes, as e.g., in semiconductor
quantum dots\cite{Eble2009}. An increase of $\tau_s$ in
the low density regime due to an admixture with unpolarized
electrons originating from an unintended background doping can be
excluded since the initial degree of polarization is constant
within the error bars for densities from $1.7 \times 10^{7}$ to
$1.7\times10^{11}$~cm$^{-2}$, i.e., for all densities where
bleaching can be neglected. The density dependence of $\tau_s$ at
10~K not only traces the transition from localized to free
carriers but also does not converge to the same $\tau_s$ at
$10^{12}$~cm$^{-2}$ as at all other temperatures. This difference
can not result from localization but reveals the different
strength of the BAP spin relaxation mechanism for
Boltzmann and Fermi statistics.

In conclusion, we have performed time- and polarization-resolved
PL measurements with a high efficient spin injection structure
that enables us to measure the spin relaxation in a (110)-GaAs-QW
over a wide range of densities. The special growth direction
provides access to exciton induced electron spin dynamics in an
electron, hole, exciton plasma that is for other growth direction
masked by the highly efficient DP spin relaxation
mechanism. The spin dynamics reveals unambiguously the existence
of excitons and proves the decrease of the exciton fraction in an
electron-hole plasma at thermal equilibrium with decreasing densities
and increasing temperatures. Although the electron, hole, exciton
plasma is well described by the Saha equation, the
currently inherent disorder in (110)-QWs and the complex spin
physics adds two degrees of complexity. We demonstrated
qualitative measurements of the exciton population in thermal
equilibrium whereat future calculations by semiconductor spin
Bloch equations, which include the commonly neglected exciton
population, will enable the extraction of the quantitative exciton
population and the influence of localization. We have additionally
verified that the BAP spin relaxation mechanism dominates at high
densities and more importantly that the dependence of the BAP
mechanism on temperature is extremely weak.

We thank W.~W.~R{\"u}hle for helpful discussions. This work has been supported by the German Science Foundation
(DFG--Priority Program 1285 ``Semiconductor Spintronics'') and the excellence cluster QUEST at the university Hannover.


\begin{thebibliography}{10}
\bibitem{note1}
The degree of PL polarization is less than 50~\% for
non-resonant excitation besides 50~\% electron spin polarization
since hot $\mathbf{k} \neq 0$ HHs have an admixture of
LHs.
\end{thebibliography}
\end{document}